\documentclass[
    superscriptaddress,
    reprint,
    showpacs,
    amsmath,
    amssymb,
    aps,
    prb,
    floatfix
]{revtex4-1}
\usepackage{physics}
\usepackage{siunitx}
\usepackage{graphicx}
\usepackage{dcolumn}
\usepackage{bm}
\usepackage{braket}
\usepackage[caption=false]{subfig}
\usepackage{float}
\usepackage{booktabs}
\usepackage{multirow}
\usepackage{tabularx}
\usepackage{adjustbox}
\usepackage{hyperref}
\usepackage{tikz}
\usepackage{makecell}

\hypersetup{
    colorlinks=true,
    linkcolor=blue,
    citecolor=blue,
    urlcolor=blue
}
\begin{document}
\title{Highly tunable optics  across a topological transition in organic polymers}
\author{D. Romanin}
\email{romanin@insp.jussieu.fr}
\affiliation{
    Sorbonne Universit\'e, CNRS, Institut des Nanosciences de Paris, UMR7588, F-75252, Paris, France
}
\author{M. Calandra}
\email{m.calandrabuonaura@unitn.it}
\affiliation{
    Department of Physics, University of Trento, Via Sommarive 14, 38123 Povo, Italy
}
\affiliation{
    Sorbonne Universit\'e, CNRS, Institut des Nanosciences de Paris, UMR7588, F-75252, Paris, France
}
\author{A. W. Chin}
\email{chin@insp.jussieu.fr}
\affiliation{
    Sorbonne Universit\'e, CNRS, Institut des Nanosciences de Paris, UMR7588, F-75252, Paris, France
}
\begin{abstract}
Controllable topological phase transitions are appealing as they allow for tunable single particle electronic properties.  
Here, by using state-of-the-art
manybody perturbation theory techniques, we show that the topological $Z_2$ phase transition occurring in the single particle spectrum of the recently synthetized
ethynylene-bridged polyacene polymers 
  is accompanied by a topological excitonic phase transition: the band inversion  in the non-trivial phase yields real-space exciton
wave functions in which electrons and holes exchange orbital characters with respect to the trivial phase.
The topological excitonic phase transition results in a broad tunability of the  singlet-triplet splittings, opening appealing perspectives for the occurrence of singlet fission. Finally, the flatness of the single-particle electronic structure in the topological non trivial phase leads to negatively dispersing triplet excitons in a large portion of the Brillouin zone, opening a route for spontaneously coherent energy transport at room temperature.  
\end{abstract}
\date{\today}
\maketitle

Pi-conjugated organic semiconductors form a highly versatile class of materials with tremendous promise for the development of cheap, flexible and non-toxic optoelectronic devices such as photovoltaics, sensors and solid state lighting \cite{kohler2015electronic,bronstein2020role}. However, bulk organic materials - typically blends of polymers and small molecules - are often highly amorphous and suffer from strong electronic disorder that limits their performance as light-harvesting materials \cite{bredas2017photovoltaic}. Recently, a striking solution for this has appeared in the form of 'on-surface' synthesis techniques which allow molecular building blocks and their emergent solid state properties to be rationally designed, fabricated and characterized for functional material applications \cite{li2021surface,sun2020surface,chen2020graphene}. 

An exemplar of such low-disorder, self-assembling materials is the family of graphene nanoribbons, whose widths, shapes, atomic compositions and electronic band structures can be controlled at a molecular scale\cite{chen2020graphene}. Remarkably, this control has also been shown to extend to the \emph{topology} of the single-particle electronic structure, leading to new emerging phenomena such as robust edge states, band inversion and spin-chain physics \cite{cao2017topological,li2021topological,zhao2021topological}. Band topology is now considered to be as fundamental as band structure and filling for future functional materials \cite{vanderbilt2018berry,breunig2021opportunities}, and the tunability of topological properties by applied fields or external conditions is a highly active area of interdisciplinary research \cite{zhao2021topological}. 

Recently, a controllable topological ($Z_2$) \emph{phase transition} has been demonstrated by varying the monomer size and chain length in a series of ethynylene-bridged polyacene polymers deposed on Au(111) \cite{CireraNatNanotec2020,GonzalezCondMat2021}. Polyacenes, such as tetracene and pentacene, are widely used in organic optoelectronics \cite{ratner2013brief}, and have attracted widespread attention in the context of next-generation photovoltaics due to their support of singlet fission (SF). SF is a spin-allowed process in which a singlet exciton decays spontaneously into a \emph{pair} of triplet excitons, allowing for enhanced photovoltaic efficiency via multiple carrier generation \cite{smith2010singlet,smith2013recent,lee2013singlet,xia2017singlet}. The impact of  topology on the excitonic properties of acene-based light harvesting structures has, to date, remained unexplored and unmeasured.

\begin{figure}[t]
\centering
 \includegraphics[width=0.9\columnwidth]{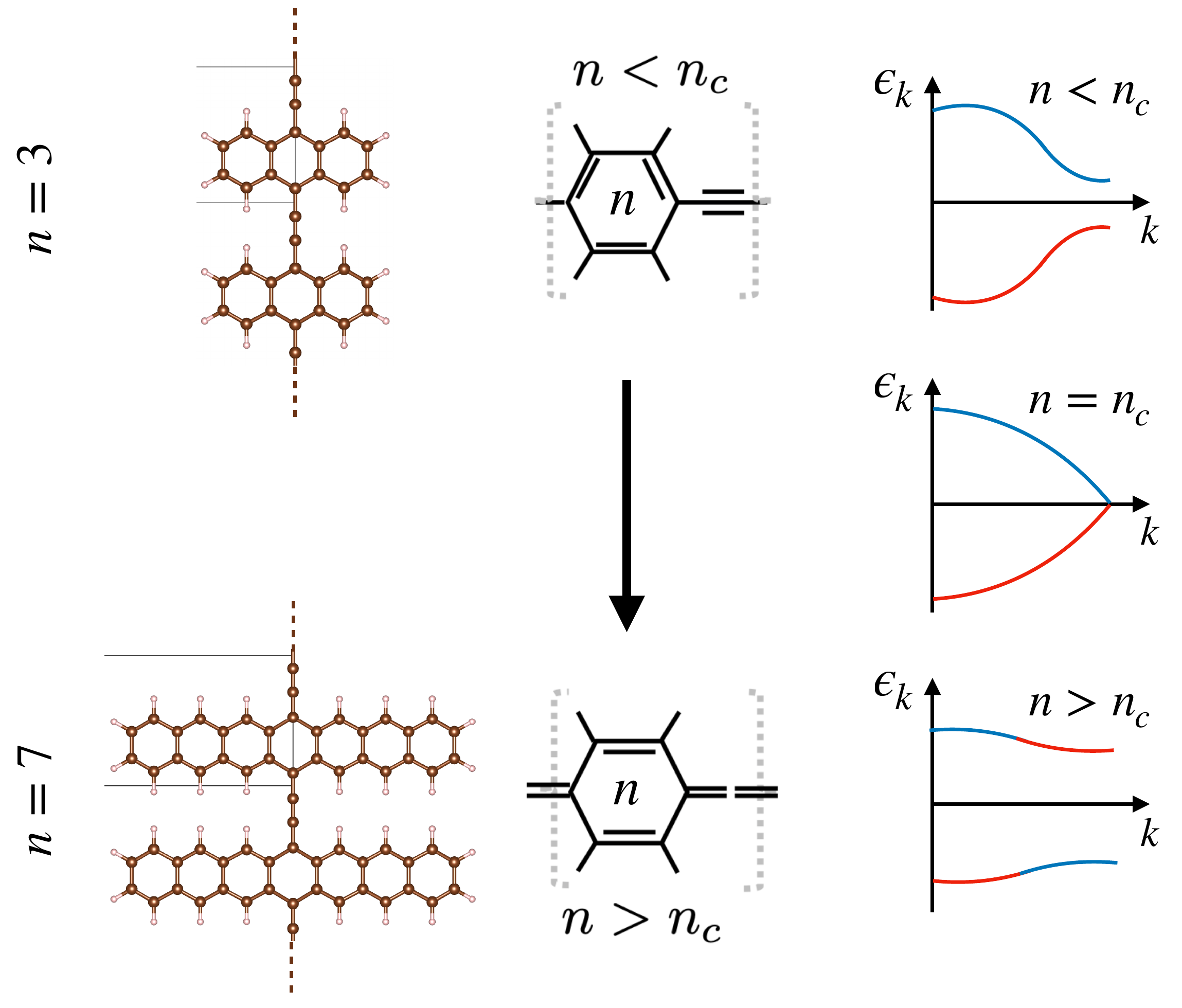}
 \caption
 {Structure of ethynylene-bridged polyacene polymers as a function of the acene length $n$. A generic polymer has the composition $(C_{4n+4}H_{2n+2})_x$, with $x$ the number of repeats. Repeats are connected by bridges that are a mixture of two resonant forms. As the $n$ increases, the dominant configuration switches from the ethylynic ($-C\equiv C-$) to a cumulenic ($=C=C=$) form, while the central ring switches from an aromatic to quinoidal form. Above a critical length $n_c$, the topologically non-trivial phase (see text) emerges, showing a smooth change of orbital character (red,blue) across the Brillouin zone (band inversion) that is absent for $n<n_c$. }
\label{fig:Fig1}
\end{figure}

In this letter, we elucidate the excitonic properties of ethynylene-bridged polyacene polymers in vacuum, and directly correlate the transition between trivial and non-trivial topological phases with dramatic changes in optical response. In particular, by using state-of-the-art electronic structure (i.e. Density Functional Theory~\cite{DovesiIJQC2014,DovesiIJQC2018} and the CAMB3LYP hybrid functional~\cite{YanaiCPL2004,TawadaJCP2004}) and many body perturbation theory (i.e. self-consistent GW approximation\cite{MariniCPC2009,Sangalli2019,FaberJCP2013} and the Bethe-Salpeter formalism~\cite{StrinatiRNC1988,BussiPS2003}), we show that the topological $Z_2$ phase transition occurring in the single particle spectrum is accompanied by a topological excitonic transition in which electrons and holes in the non-trivial phase exchange orbital character with respect to the 
trivial one. The large tunability of the energy separation between 
singlet and triplet excitons as a function of the monomer length in the trivial phase opens promising perspectives for singlet fission in this material. Finally, the flatness of the single particle electronic structure in the non-trivial phase leads to a  \emph{negative} dispersion for triplet excitons in a large portion of the Brillouin zone, opening a route for spontaneously coherent energy transport at room temperature.
 Numerical details can be found in the Supplemental Material (SM, Sec.~S1). 

Figure \ref{fig:Fig1} shows the structure of ethynylene-bridged polyacene polymers. Each polymer is composed by polyacene monomers
connected by ethynylene-bridges. The number of aromatic rings in each monomer is labeled $n$. The structure can be made long at will in the bridge direction (one dimensional periodicity).
Both anthracene ($n=3$) and pentacene ($n=5$) polymers have recently been synthesized. Usually, the HOMO-LUMO gap of acene monomers decreases gradually as $n$ increases, and tends to a finite limit as $n\rightarrow\infty$ \cite{andre2001linear}. However, it was observed in Refs. \cite{CireraNatNanotec2020,GonzalezCondMat2021} that the band gap in the acene polymers on Au(111) almost closes completely when going from $n=3$ (anthracene) to $n=5$ (pentacene), and this was accompanied by a marked change in the bond-length alternation (BLA) of the polymer bridges as well as a transition from an aromatic to quinoidal configuration for the central carbon ring in the acene monomer (Fig. \ref{fig:Fig1}). In Ref. \cite{CireraNatNanotec2020}, the former observation was rationalized by a mapping of the electronic structure of the polymers to the  Su-Schrieffer-Heeger (SSH) model \cite{vanderbilt2018berry} and to the interplay between the 
 intra and inter (bridge-mediated) monomer electron hopping amplitudes that can be tuned by changing $n$.
 The topological nature of the acene polymers is confirmed by the dection of  localized 'edge states' in sufficiently long-but-finite pentacene ($n=5$) polymers\cite{CireraNatNanotec2020,GonzalezCondMat2021}.
\begin{table}[t]
\begin{tabular}{ c c c c c c c c}
 \hline
 \hline
 n & c [$\AA$] & BLA [$\AA$] &  $\Delta E$  & $\Delta E$ & S$_I$  & E$_B$ & S$_I$-T$_I$ \\
   & {\scriptsize (Hybrid)} &  {\scriptsize (Hybrid)}           &   {\scriptsize (Hybrid)}                & {\scriptsize (evGW)}           &   {\scriptsize (BS)}      &   {\scriptsize (BS)}       & {\scriptsize (BS)}  \\
 \hline
 3 & 6.89 & 0.215 &  3.48 & 3.33 & 1.84 & 1.49 & 0.49 \\
 5 & 6.84 & 0.210 &  2.35 & 1.53 & 0.77 & 0.76 & 0.30 \\
 7 & 6.91 & 0.098 &  3.31 & 2.78 & 1.66 & 1.12 & 0.62 \\
 \hline
 \hline
\end{tabular}
\caption{Lattice parameter (c), bond length alternation (BLA),  quasiparticle bandgap ($\Delta E$), first singlet energy (S$_I$) and binding energy ($E_B$), and first singlet-triplet splitting (S$_I$-T$_I$) for ethynylene-bridged poly-anthracene (n=3), poly-pentacene (n=5) and poly-heptacene (n=7) in vacuum. Hybrid refers to the CAMB3LYP functional~\cite{YanaiCPL2004,TawadaJCP2004} and evGW to the GW method with self-consistency on the eigenvalues. The excitonic energies are calculated with the Bethe-Salpeter (BS) approximation.}
\label{table:1}
\end{table}

\begin{figure}
\centering
 \includegraphics[width=1.0\linewidth]{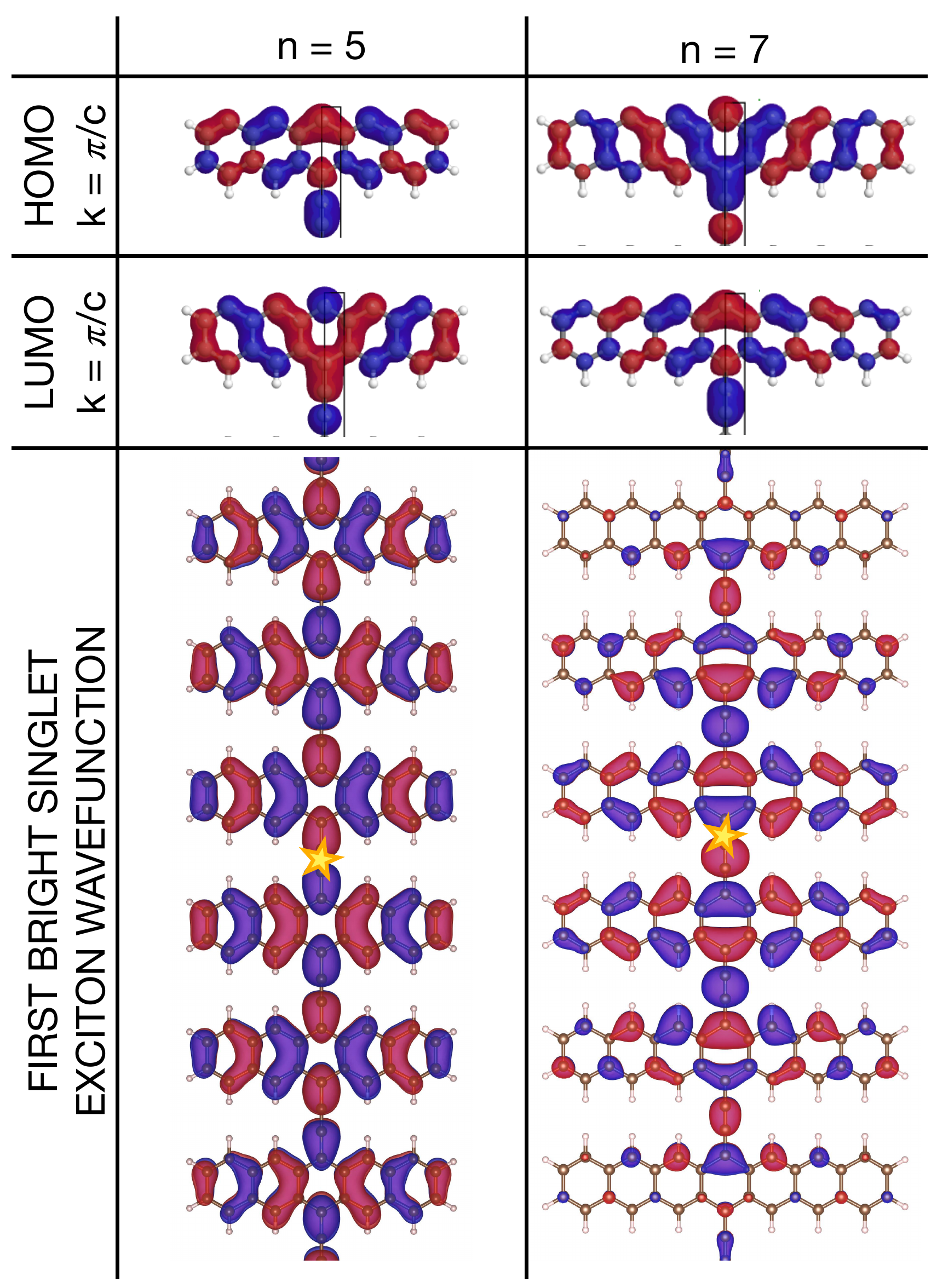}
 \caption
 {\textit{Upper panels:} band-edge ($k=\pi/c$) HOMO/LUMO single-particle wave functions (with relative phases) for $n=5$ and $n=7$. \textit{Lower panel:} exciton wavefunction (with relative phase) of the first bright singlet ($S_I$, see Fig.~\ref{fig:Fig3}) for $n=5$ and $n=7$ at $q=0$. The yellow star fixes the position of the hole.
   }
\label{fig:Fig2}
\end{figure}

\begin{figure*}[t]
\centering
 \includegraphics[width=0.75\linewidth]{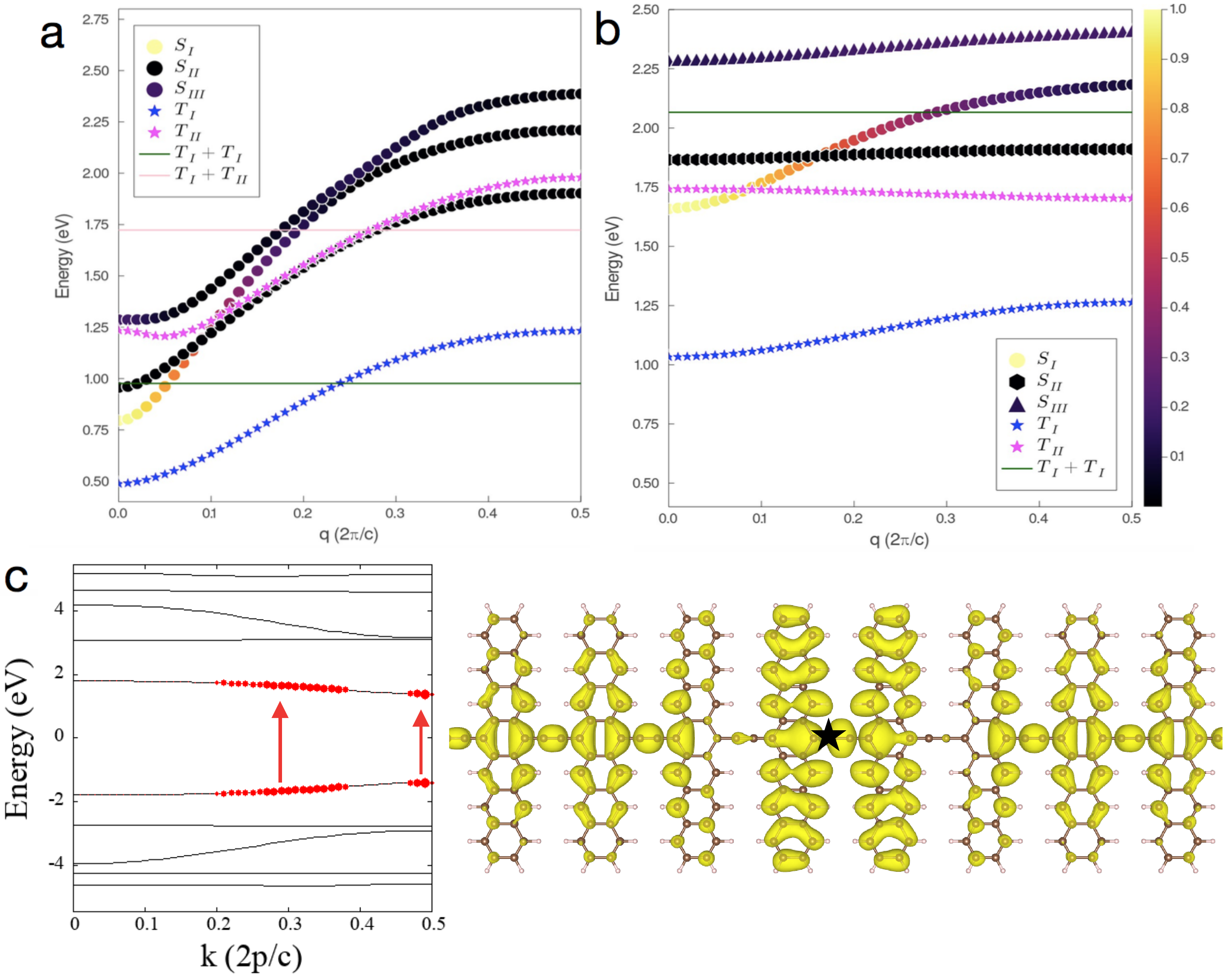}
 \caption
 {Exciton dispersions for $n=5$ (a) and $n=7$ (b) showing the energy threshold for triplet pair creation (horizontal lines). Singlet exciton dispersions are coloured according to their normalized optical oscillator strength (color bar), with the brightest state having unit strength. (c) Single-particle transitions responsible for the $S_{III}$ singlet excitonic state at $q=0$ for poly-heptacene ($n=7$), together with the modulus squared of  its exciton wave function. The hole is localized at black star. This state shows a striking change in the electronic orbitals away from the hole (indicated by a blue circle), which is a consequence of the topological band inversion for $n=7$.}
\label{fig:Fig3}
\end{figure*}

  Tab.~\ref{table:1} reports the result of the 
  single-particle band-gap, as well as the theoretical values we have obtained via DFT (CAMB3LYP) and DFT$+$many-body perturbation theory (self-consistent GW on-eigenvalues-only, i.e. evGW) for
  $n=3,5,7$ in vacuum. We consider here a paramagnetic ground state. The possible occurrence of magnetic states at very low temperature is discussed in Sec.~S7 in the SM. All polymers show a direct single-particle band-gap at the edge ($k=\pi/c$) of the Brillouin zone (see Fig.~S4-S5 in SM). The gaps decrease from $n=3$ to $n=5$, then \emph{increase} for $n=7$, supporting a picture of a topological phase transition (band closure) between pentacene and heptacene. We also note that while CAMB3LYP and evGW give very similar gaps for $n=3$,  the evGW band gap becomes smaller with respect to CAMB3LYP for $n=5,7$. This suggests the increasing role of electronic correlations as we approach and traverse the transition: approximate functionals cannot fully take these into account, so more accurate many-body techniques yield larger corrections. The largest deviations occurs for $n=5$ ($\approx 39\%$), which is the closest system to the topological transition (smallest QP gap). The complete band structures computed at different levels of theory for $n=3-7$ are presented in the SM (Sec.~S3).

In Fig.~\ref{fig:Fig2} we show the CAMB3LYP valence ($v$), or HOMO, and conduction ($c$), or LUMO, wave functions at the band edges ($k=\pi/c$) for $n=5$ and $n=7$. Although there is a perturbation from the bridge states, for poly-pentacene these orbitals are similar to the HOMO (LUMO) molecular orbitals of the $n=5$ monomer. The same is also seen for $n=3$ (see SM, Sec.~S4). However, for heptacene ($n=7$), the $v$ and $c$ states correspond, respectively, to the LUMO and HOMO orbitals, i.e. the $v$ and $c$ states have {\it inverted} orbital character, as is expected in the SSH model. Moreover, this inversion should occur smoothly across the band structure, as it is verified in Figs. S6 in the SM by comparing the HOMO (LUMO) orbital of $n=7$ at the center of the Brillouin zone ($k=0$) with the LUMO (HOMO) orbital at the border of the Brillouin zone ($k=\pi/c$) \footnote{Modifications of the HOMO (LUMO) monomer orbitals due to the pi orbitals of the bridge are present in all cases}. No such orbital inversion across the band structure is seen for $n=3,5$. Thus, in vacuum, the topological phase transition occurs for monomers lengths between $n=5$ (trivial) and $n=7$ (nontrivial). This is further supported by the behaviour of the BLA of the bridge atoms, as shown in Tab \ref{table:1}. On going from $n=5\rightarrow n=7$, the BLA dramatically reduces, indicating a switch from alternating single and triple bonds to a cumulenic double-double bond resonance form with nearly equal bond lengths. Complete structural characterization can be found in the SM (Sec.~S2).

\begin{figure*}[t]
\centering
 \includegraphics[width=1.0\linewidth]{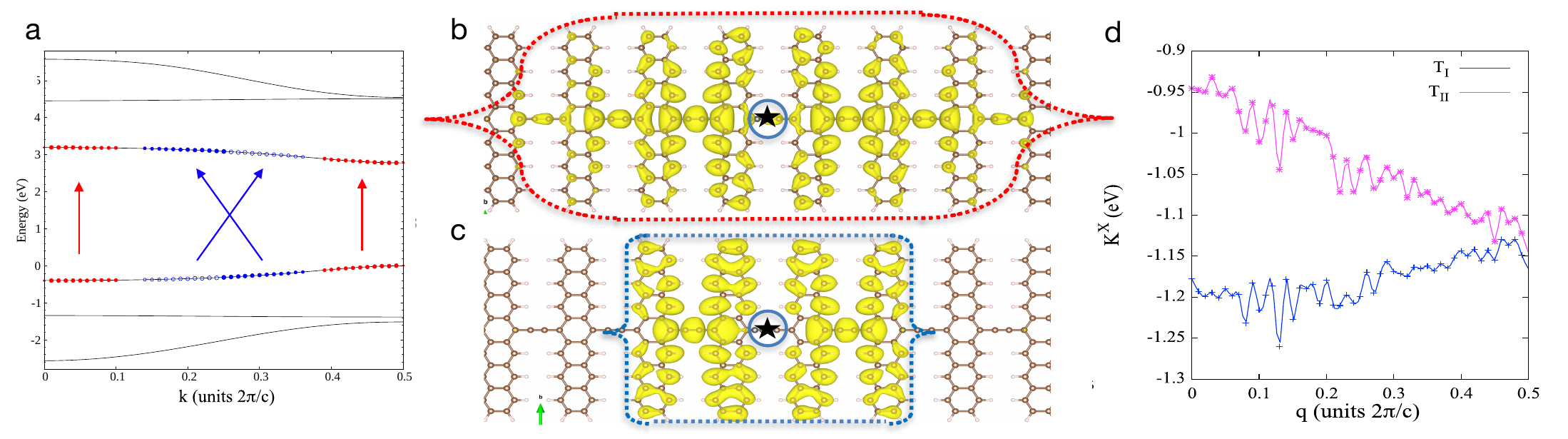}
 \caption
 {Poly-heptacene (n=7): (a) single-particle transitions responsible for $T_{II}$ at $q=0$ (red) and $q=\pi/c$ (blue);(b-c) Modulus squared of the exciton wave functions for $T_{II}$ at $q=0$ (b) and $q=\pi/c$ showing contraction of the exciton at finite momentum, as well as qualitative change in electron density due to band inversion. Holes are localized at circled position. (d) Screened direct Coulomb interaction $K^X$ averaged over triplet eigenfunctions as a function of $q$, obtained with Yambopy.}
\label{fig:Fig4}
\end{figure*}

Exciton states and dispersion were obtained by solving the Bethe-Salpeter equation (BSE) using the evGW quasi-particle energies. 
The excitation and binding energies of the lowest (optically bright) singlet and triplet excitons are given in Tab \ref{table:1} for $n=3,5,7$. The excitonic gaps follow the same trend as the band structure, as do the large excitonic binding energies ($\sim0.3-0.6$ eV). Due to strong coulomb interactions and low dimensionality, a large exchange splitting between singlet triplet excitons ($\sim0.3/0.6$ eV) is always present (Tab.~\ref{table:1}). We note that the topological transition between $n=5$ and $n=7$, has dramatic effects on the manybody properties of the polymers as: (i) the optical gap is dramatically reduced ($0.77$ eV for $n=5$, c.f. $1.8$ eV in molecular crystals), (ii) the reduction of the binding energy of the first singlet state (see Tab. \ref{table:1}) for $n=5$ implies a complete delocalization of the excitonic wavefunction with a transition from Frenkel to Mott-Wannier excitons across the topological transition ($n=5$ to $n=7$). The change in localization of the excitonic wavefunction is clearly shown in Fig. \ref{fig:Fig2} (bottom panel) where, crucially, the change in character of the excitonic wavefunction is also evident . Indeed, for $n=5$ the excitonic wavefunction is mainly built from the LUMO state, while for $n=7$ the HOMO becomes dominant, a fingerprint of the topological transition on the manybody states. The extreme tunability of the optical properties across the topological transition suggests that these polymers are an ideal platform for singlet fission as  we now discuss in the context of the excitonic dispersions.    

 Fig.~\ref{fig:Fig3} compares the exciton dispersions $E^\lambda (q)$ for $n=5$ and $n=7$. Singlet excitons are optically spin-allowed, and have been plotted to take into account their \textit{optical strength}, i.e. $ f_\lambda(q)=\abs{\sum_{c,v,k}A_{c,v}^{\lambda}(k,q)\bra{c,k}\mathbf{D}\ket{v,k}}^{2}$,
where $\ket{ck}$ ($\ket{vk}$) is the Bloch wave function for an electron (hole) with momentum $k$ on band $c$ ($v$), $\mathbf{D}$ is the electric dipole operator (here, aligned along polymer repeat axis)
and $A_{c,v}^{\lambda}(k,q)$ is the amplitude of a given electron-hole transition in the excitonic ($\lambda$) wave function in $k$ space.

For $n=5$, we find a rather congested spectrum containing both bright and dark singlet excitons, as well as two triplet bands. Notably, the lowest bright singlet exciton ($S_I$) is extraordinarily dispersive and remains optically active over more than $60\%$ of its bandwidth ($1.6$ eV). After this it's remanant oscillator strength is donated to the less dispersive $S_{III}$ state (optically active at $q=0)$ through an avoided crossing. This is important for singlet fission (SF) as, although the $q=0$ $S_I$ state lies below the threshold energy for SF ($\approx 2\times E_{T_I}$, as shown on Fig.\ref{fig:Fig3}), its rapid dispersion creates a large, energetically alllowed phase space for SF at finite momentum. Finite-$q$ SF has previously been shown to be an effective mechanism in pentacene ($n=5$) thin films \cite{refaely2017origins}, even though the joint phase space is much smaller, due to the narrow dispersion of singlet excitons in films \cite{refaely2017origins}. Strikingly, the $n=5$ polymer also permits ( on energetic grounds) the decay of $S_I$ at large-$q$ into an excited $T_I T_{II}$ triplet pair (Fig.\ref{fig:Fig3}). Recently, Pandya \textit{et al}. have shown that optical excitation of a $(T_I T_I)\rightarrow (T_I T_n)$ transition induces spatial separation of the tightly bound triplet pairs that form the lowest excited state of blue polydiacetylenes, suggesting that the new SF pathway available in the $n=5$ polymer could also spontaneously generate 'free' triplets pairs that preserve their germinate spin entanglement across real space \cite{pandya2020optical}.          

Another critical aspect of SF is the matrix element for the conversion of bright singlets into triplet excitons. In acenes, this is arises via a super-exchange mechanism involving a higher-lying 'charge transfer' singlet exciton \cite{berkelbach2013microscopic}. This matrix element takes the generic form $\chi_{SF}=(M_{S_I}^{S_n}M_{S_n}^{TT})(\Delta E_{S_I S_n})^{-1}$, where $M_{x}^y$ is a Coulombic transition amplitude between states $x$ and $y$ that are separated in energy by $\Delta E_{xy}$. As shown in Fig.\ref{fig:Fig3}a, the $S_{III}$ state  - which has strong charge-transfer character (see SM Sec.~S6 for a complite visualization of the excitonic wavefunctions) - mixes strongly  with $S_I$ near the avoided crossing ($\Delta E_{S_I S_{III}}\approx 0)$. Consequently, a large super-exchange $\chi_{SF}$ could be also be expected above threshold, leading to concomitantly large SF rates. We also note that $S_{III}$ at $q=0$ is already above threshold and is optically bright, permitting \emph{direct} SF from this CT state \cite{berkelbach2013microscopic}.                    

On the other side of the topological transition ($n=7$), the exciton dispersions become significantly flatter, matching the underlying band structure (cf. Fig.\ref{fig:Fig3}b and Fig.~S5c). The larger relative energy of the $T_I$ states reduces the phase space for finite-$q$ SF, although the $S_{III}$ band remains above threshold for all momenta. 
The key change of relevance for SF is the effect of band inversion on the SF matrix elements. As seen in Fig.\ref{fig:Fig3}c, the square modulus of the excitonic wavefunction undergoes a stark evolution from a HOMO to LUMO-like distribution, as the $e-h$ separation increases. This arises due to the strong e-h localisation in real space: the corresponding $k$-space wave function contains e-h transitions that extends across region of band inversion, i.e. single particle state from the non-inversion to the inversion region of the HOMO/LUMO states.

 Perhaps the most striking feature in Fig. \ref{fig:Fig3} is the existence of \emph{negatively} dispersing $T_{II}$ excitons. For $n=5$ this negative dispersion is confined to a small region around $q=0$; for $n=7$ it spectacularly extends across the entire band.  Indeed, $T_{II}$ decreases across the band by $\approx 40$ meV ($K_B T \approx 25$ meV at $298$K), so that phonon scattering should \emph{spontaneously} relax these excitons into stable, large-momentum propagating states. During the $T_{II}\rightarrow T_I$ lifetime, and at low temperatures, this could provide a rather novel, coherent energy transport mechanism in which energy relaxation actually {\it promotes} ballistic motion over diffusion. As previously noted, the $n=5$ polymer also supports SF into $T_I T_{II}$ pairs, in which the negative dispersion of the $T_{II}$ state could additionally promote triplet separation.          

With a view to understanding and controlling these anomalous dispersions, Fig.\ref{fig:Fig4}a compares the $k$-space distributions of e-h transitions in the $T_{II}$ state at $q=0$ and $q=\pi/c$ on the $n=7$ quasiparticle band structure. At $q=0$, the $T_{II}$ state is built by e-h pairs clustered around the band edge and zone center, and at $q=\pi/c$ the dominant transitions lies in the centre of the band. According to the effective Bethe-Salpeter Hamiltonian, the energy of the triplet exciton states contains two contributions: the (non-interacting) kinetic energy of the e-h quasiparticles and the matrix element between the conduction $c$ and valence $v$ band of the Fourier transform of the screened direct Coulomb interaction $K^X_{\substack{cvk\\c'v'k'}}$ at k-points $k$ and $k'=k+q$. Due to the flatness of the band structure in Fig.\ref{fig:Fig4}, there is essentially no change in the carrier's kinetic energy as $T_{II}(q=0)\rightarrow T_{II}(q=\pi/c)$, whereas this normally provides a strong, positive contribution to the exciton dispersion, as it does for $n=5$. 

At the same time, the direct Coulomb interaction matrix element that mixes e-h transitions in the excitonic wave function arises from a screened interaction, and this screening is $q$-dependent. Figure \ref{fig:Fig4} shows $K^X (q) = \sum_{\substack{cvk\\c'v'k'}}(A_{c',v'}^{\lambda}(k',q))^*A_{c,v}^{\lambda}(k,q)K^X_{\substack{cvk\\c'v'k'}}$ for the $n=7$ triplet excitons, showing that $K^X (q)$ is a continuously \emph{decreasing} function of $q$ for $T_{II}$, i.e. the binding energy of the exciton becomes \emph{greater} at the band edge. Combined with the neutral change in carrier Kinetic energy (K.E.) - which is induced by the flat bands that appear above the topological transition -  we then see that the negative dispersion arises from reduced screening at short wavelengths (large $q$). This interpretation is visualized directly in Figs.\ref{fig:Fig4} b$\&$c, where the shrinkage (increased binding energy/decreasing Bohr radius) of the exciton can be seen in the real-space wave function as $q\rightarrow \pi/c$. We have also found that the negative dispersion for $n=5$ also arises from a similar effect, although rapidly rising carrier K.E. overwhelms the effect at large $q$. Thus, it is the flat bands that \emph{must} appear above the topological transition (the disconnected dimer limit, in the SSH model) that promote anomalous exciton dynamics.

In conclusion, we have shown that the topological transition occurring in  ethynylene-bridged acene polymers transfers to the excitonic properties and lead to an extremely tunability of the optical properties across the transition. The variation of the exciton binding energies across the transition leads to a transition between Frenckel and Mott-Wannier excitons, as a function of the acene length. The wide range of singlet-triplet splittings could also be enhanced by the substrate on which the polymers are grown or deposed. Indeed, the screening from the substrate will substantially reduce the effects of the unscreened, long-range vacuum interactions, allowing for additional modulation
of optical gaps, excitonic binding energies, S-T splittings and - most importantly - of the critical length $n$ of the acene for the occurrence of the topological transition.  Indeed, Cirera \textit{et al.}'s placement of $n=5$ in the non-trivial phase is likely related to the metallic substrate inducing a gap reduction (see SM, Sec.~S3). Additionally, based on the findings of previous work on carbynes \cite{RomaninJPCL2021} and finite-sized acene polymers \cite{GonzalezCondMat2021}, the intriguing change from acetylenic to cumulenic bridging at the transition suggests that phonons and anharmonicity could also play an important, temperature-dependent role in bringing these systems to the critical point where, if stabilized, an intrinsic organic metal is expected to emerge.\\

The authors acknowledge support from the ANR project ACCEPT (Grant No. ANR-19-CE24-0028). This work was granted access to the HPC resources of IDRIS, CINES and TGCC under the allocation 2021-A0100912417 made by GENCI. D. R. thanks D. Varsano and F. Paleari for fruitful dicussion on the GW+BSE computations. The data that supports the findings of this study are available from the
corresponding author on reasonable request.

\bibliography{references}

\pagebreak
\widetext
\begin{center}
\textbf{\large Supplemental Materials: Highly tunable optics across a topological transition in organic polymers}
\end{center}
\setcounter{equation}{0}
\setcounter{figure}{0}
\setcounter{table}{0}
\setcounter{page}{1}
\makeatletter
\renewcommand{\theequation}{S\arabic{equation}}
\renewcommand{\thefigure}{S\arabic{figure}}
\renewcommand{\bibnumfmt}[1]{[S#1]}
\renewcommand{\citenumfont}[1]{S#1}

\section{Computational details}

Density functional theory calculations on the groundstate geometry (i.e. unit cell and atomic positions) are performed in vacuum with localized gaussian basis, using CAMB3LYP hybrid xc functional~\cite{YanaiCPL2004,TawadaJCP2004} and CRYSTAL17~\cite{DovesiIJQC2014,DovesiIJQC2018}. The CAMB3LYP hybrid functional includes long-range corrections to the B3LYP, which is particularly suited for carbon-based materials, and has been shown to provide accurate results for charge transfer excitations, excitations to Rydberg states, polarizability of long chains and vibrational properties comparable to those obtained via Quantum Montecarlo methods~\cite{RomaninJPCL2021} for systems in vacuum. We have employed the triple-$\xi$-polarized Gaussian type basis set\cite{Vilela-OliveiraJCC2019} with real space integration tolerances of 10-10-10-25-50 and an energy tolerance of $10^{-12}$ Ha for the total energy convergence. Each polymer has its principal axis along the $x$ direction, with $\sim500$~\AA~of vacuum along the non-periodic $y$ and $z$ directions in order to have a proper one-dimensional system. The Brillouin zone is sampled with a uniform mesh of $50\times1\times1$ k-points.\\

Many-body corrections to the single-particle bandstructure are computed within the GW approximation using the Yambo~\cite{MariniCPC2009,Sangalli2019} code and the plasmon-pole approximation~\cite{RojasPRB1995}. Eigenvalues and wavefunctions are initially computed with the Perdew-Burke-Ernzerhof (PBE) exchange-correlation (xc) functional and Quantum ESPRESSO~\cite{GiannozziIOP2009,GiannozziIOP2017}. Each polymer has its principal axis along the $x$ direction, with $\sim10$~\AA~of vacuum along the non-periodic $y$ and $z$ directions as well as a cut-off on the Coulomb potential~\cite{RozziPRB2006} in order to have a proper one-dimensional system. In this case we have employed a norm-conserving pseudopotential with the kinetic energy cut-off set to $70$ Ry and a threshold of $10^{-10}$ Ry on the total energy.

We performed self-consistent $GW$ calculations on eigenvalues only (ev$GW$)~\cite{FaberJCP2013} for both the Green's function $G$ and the screened electron-electron interaction $W$ in the Plasmon-Pole approximation (PPA). Excitonic effects are then evaluated by solving the Bethe-Salpeter~\cite{StrinatiRNC1988,BussiPS2003} equation (BSE) in the Tamm-Dancoff approximation on top of the ev$GW$ band structure. The number of k-points has been fixed to 100 in order to converge the first two singlet excitonic peaks in the absorption spectrum. Finally, we want to specify that results for the excitons are obtained via a dynamical screening calculated using QP eigenvalues for the BSE equation.\\

In Tab.~\ref{table:1} we summarize the converged parameters used for a GW calculation for each polymer (identified by n). The polarization of the electric field is along the principal axis of the polymers. The labels of the parameters are those used in the Yambo software\cite{MariniCPC2009,Sangalli2019}, whose meaning is the following:
\begin{itemize}
    \item EXXRLvcs: G-vectors used in the sum for the exchange part of the self energy;
    \item BndsRnXp: Number of bands entering in the sum over states in the RPA response function;
    \item NGsBlkXp: Size of the dielectric matrix in G-space;
    \item GbndRnge: Number of bands entering in the sum over states in the correlation part of the self energy;
    \item PPAPntXp: Plasmon pole Imaginary Energy.
\end{itemize}

\begin{table}[H]
\centering\begin{tabular}{|c|c|c|c|c|c|}
 \hline
 n & EXXRLvcs [Ry] & BndsRnXp & NGsBlkXp [Ry] & GbndRnge [Ry] & PPAPntXp [eV]\\
 \hline
 3 & 100 & 100 & 5.0 & 100 & 54.42276\\
 \hline
 5 & 50 & 75 & 5.0 & 75 & 54.42276\\
 \hline
 7 & 25 & 100 & 5.0 & 100 & 54.42276\\
 \hline
\end{tabular}
\caption{GW converged parameters}
\label{table:1}
\end{table}

In Tab.~\ref{table:2} we summarize the converged parameters used for a BSE calculation for each polymer (identified by n). The polarization of the electric field is along the principal axis of the polymers. The labels of the parameters are those used in the Yambo software\cite{MariniCPC2009,Sangalli2019}, whose meaning is the following:
\begin{itemize}
    \item \# Valence/Conduction bands: band states from which the electron-hole basis of the BSE kernel is constructed;
    \item BSENGexx: G-components to be summed in the Exchange part of the BSE kernel;
    \item BSENGBlk: Number of G-vectors of the Screened Coulomb Potential matrix W(G,G'), to be included in the sum of the e-h attractive Kernel.
\end{itemize}

\begin{table}[H]
\centering\begin{tabular}{|c|c|c|c|c|}
 \hline
 n & \# Valence bands & \# Conduction bands & BSENGexx [Ry] & BSENGBlk [Ry]\\
 \hline
 3 & 5 & 5 & 10 & 1\\
 \hline
 5 & 3 & 4 & 20 & 3\\
 \hline
 7 & 4 & 4 & 100 & 5\\
 \hline
\end{tabular}
\caption{BSE converged parameters}
\label{table:2}
\end{table}

In each case, we employed a truncated Coulomb cutoff~\cite{RozziPRB2006} in order to correctly represent a 1D system.

\newpage

\section{Structural properties}

\begin{figure}[H]
 \centering\includegraphics[height=0.25\textheight,keepaspectratio]{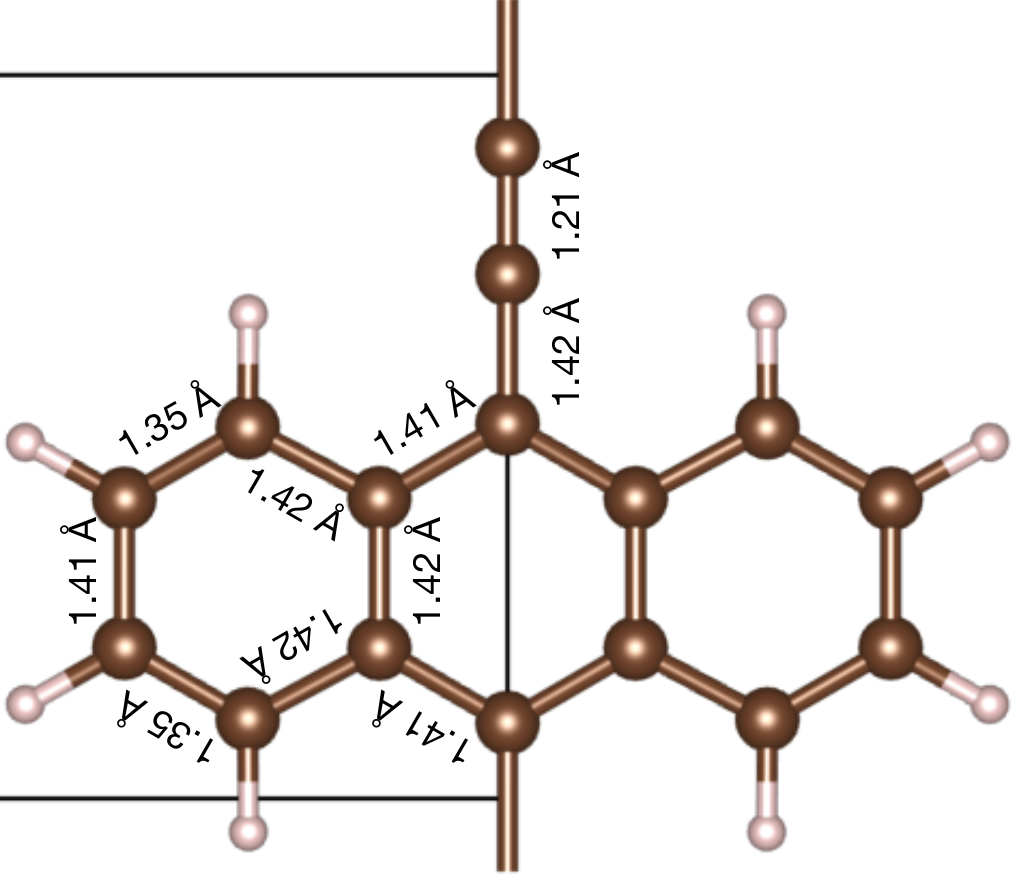}
 \caption
 {Stick-and-ball model of poly-anthracene (n=3) with relative bond lengths. The groundstate geometry has been obtained via the CAMB3LYP~\cite{YanaiCPL2004,TawadaJCP2004} hybrid functional. The black horizontal lines indicate the unit cell. Bond-lengths are written only for the left part of the polymer since it is symmetric with respect to its reciprocal axis.
   }
\label{fig:structure_n3}
\end{figure}

\begin{figure}[H]
 \centering\includegraphics[height=0.25\textheight,keepaspectratio]{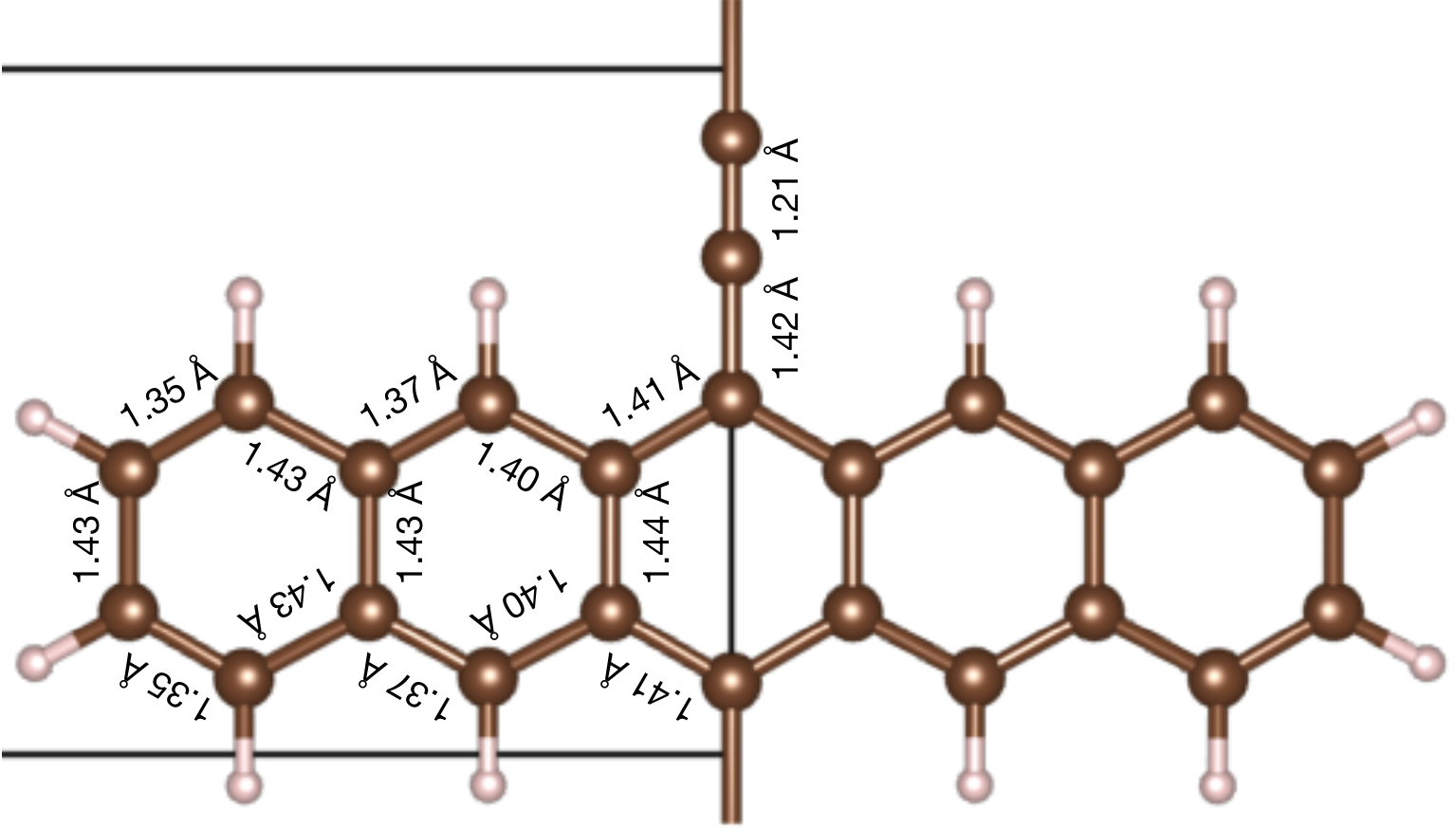}
 \caption
 {Stick-and-ball model of poly-pentacene (n=5) with relative bond lengths. The groundstate geometry has been obtained via the CAMB3LYP~\cite{YanaiCPL2004,TawadaJCP2004} hybrid functional. The black horizontal lines indicate the unit cell. Bond-lengths are written only for the left part of the polymer since it is symmetric with respect to its reciprocal axis.
   }
\label{fig:structure_n5}
\end{figure}

\begin{figure}[H]
 \centering\includegraphics[height=0.25\textheight,keepaspectratio]{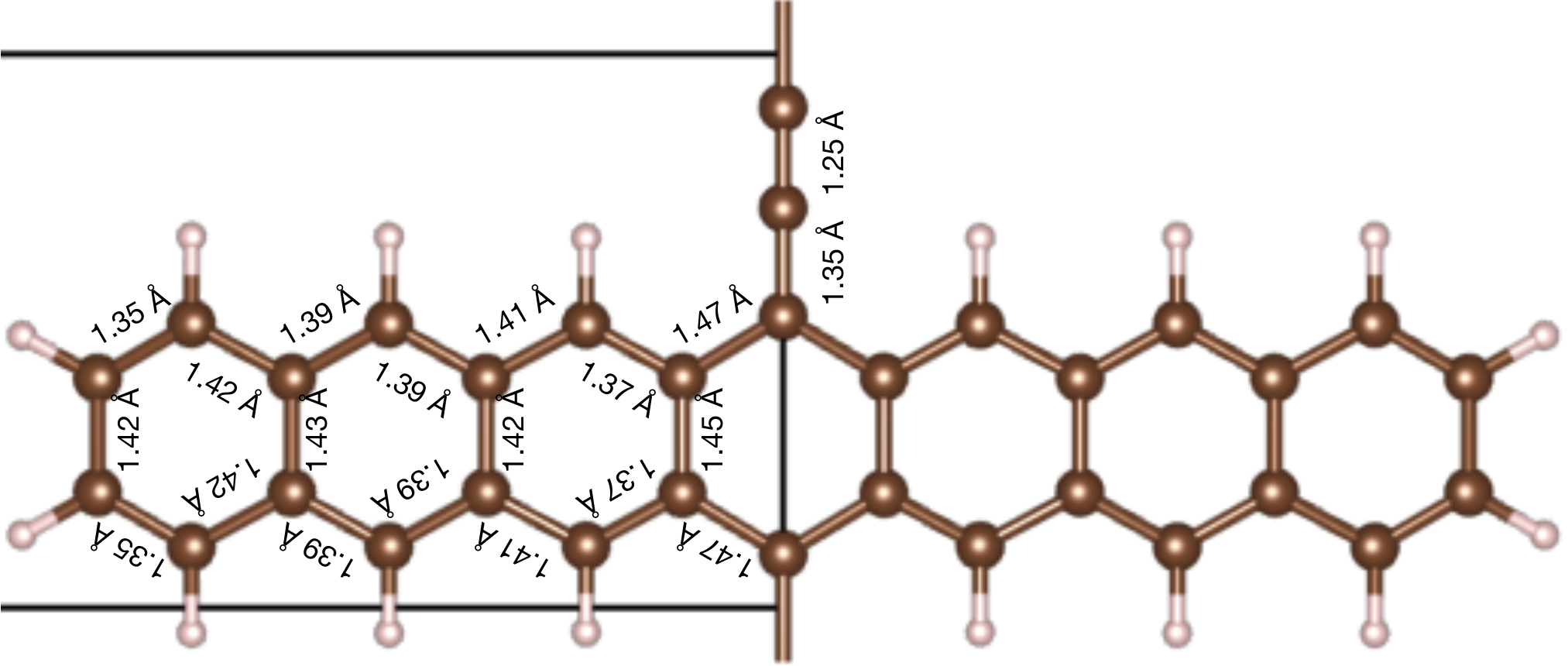}
 \caption
 {Stick-and-ball model of poly-heptacene (n=7) with relative bond lengths. The groundstate geometry has been obtained via the CAMB3LYP~\cite{YanaiCPL2004,TawadaJCP2004} hybrid functional. The black horizontal lines indicate the unit cell. Bond-lengths are written only for the left part of the polymer since it is symmetric with respect to its reciprocal axis.
   }
\label{fig:structure_n7}
\end{figure}

\section{Electronic bandstructure}

\begin{figure}[H]
 \centering\includegraphics[width=1.00\linewidth]{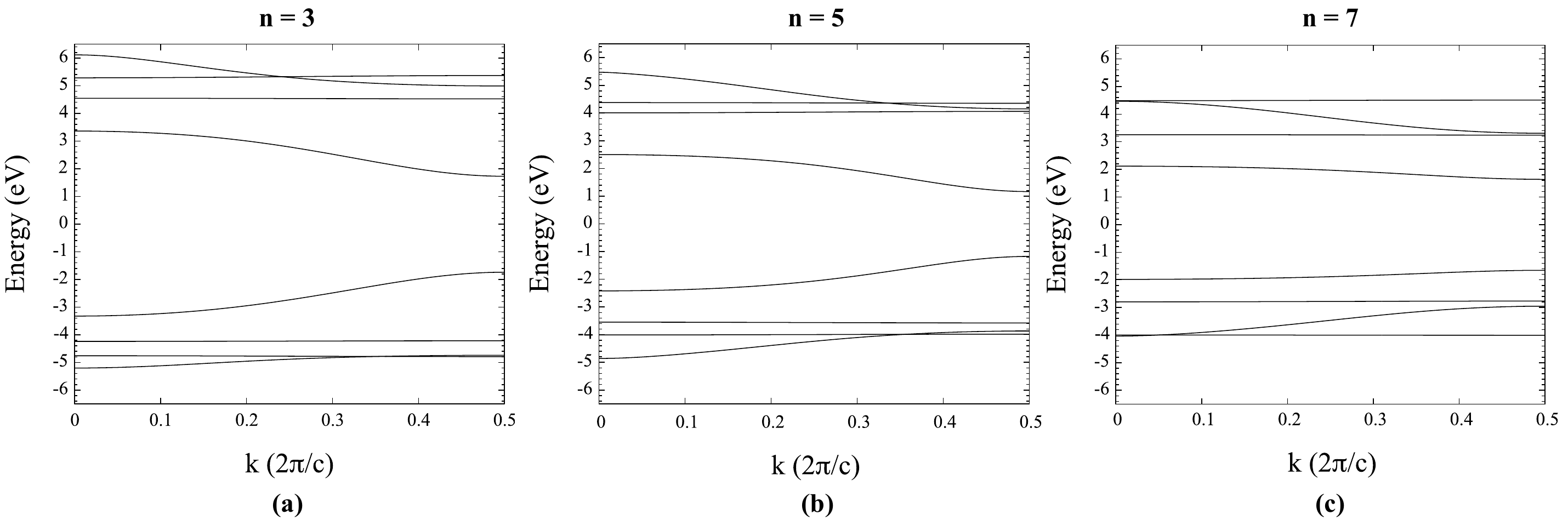}
 \caption
 {Electronic bandstructure for poly-anthracene (n=3), poly-pentacene (n=5) and poly-heptacene (n=7) obtained via the CAMB3LYP~\cite{YanaiCPL2004,TawadaJCP2004} hybrid functional.
   }
\label{fig:CAMB3LYP_bands}
\end{figure}

\begin{figure}[H]
 \centering\includegraphics[width=1.0\linewidth]{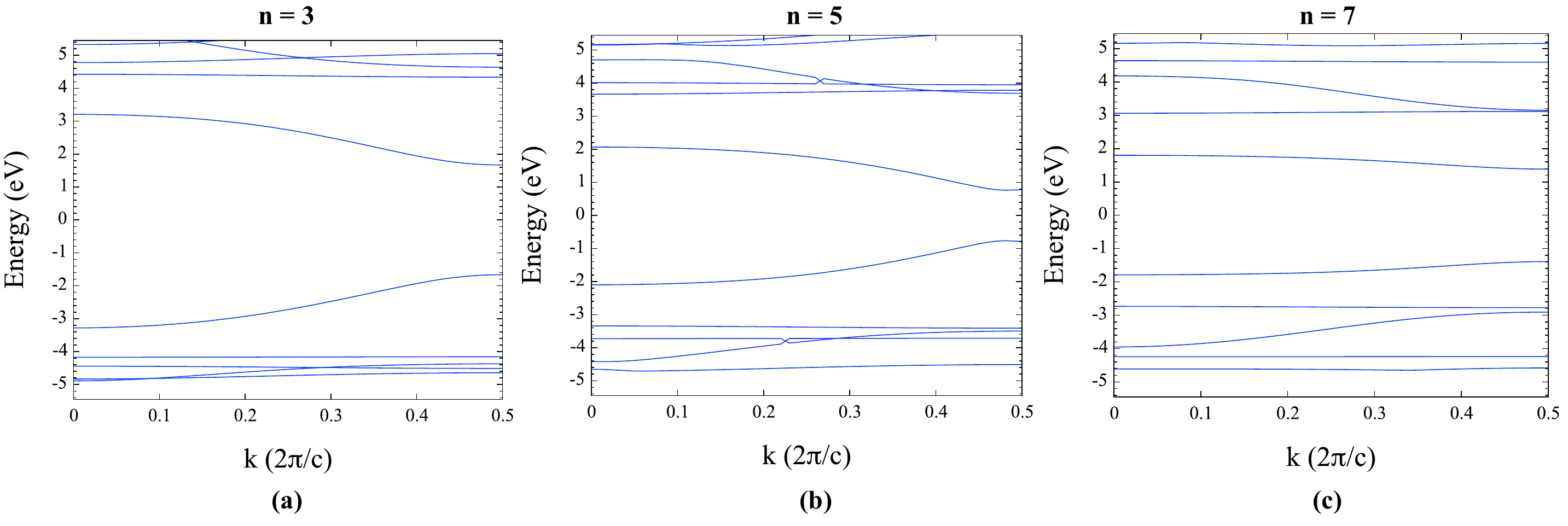}
 \caption
 {Electronic bandstructure for poly-anthracene (n=3), poly-pentacene (n=5) and poly-heptacene (n=7) obtained via self-consistent GW.
   }
\label{fig:evGW_bands}
\end{figure}

\section{HOMO/LUMO wavefunctions}

\begin{figure}[H]
 \centering\includegraphics[width=1.0\linewidth]{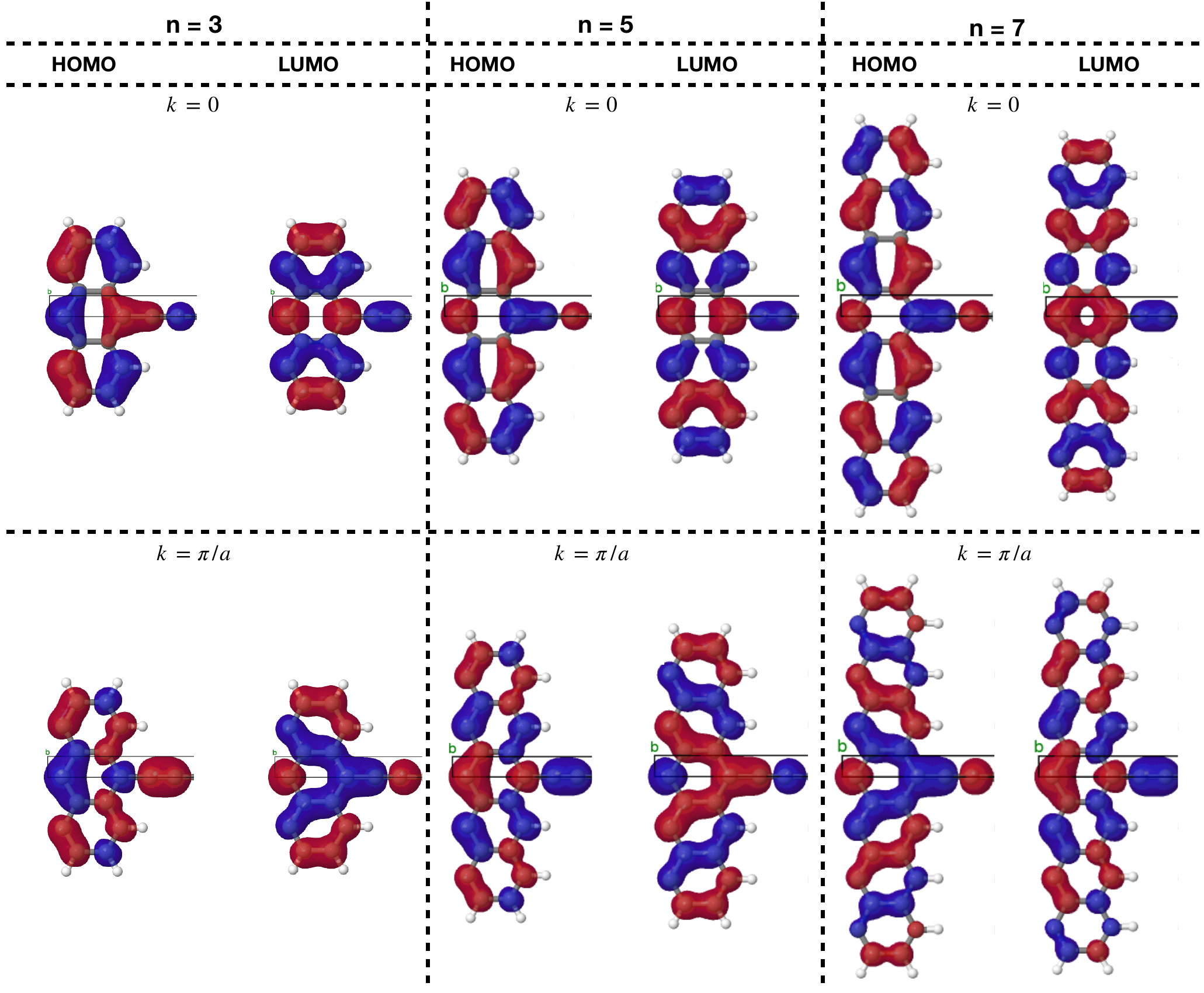}
 \caption
 {HOMO/LUMO wavefunctions at the center ($k=0$) and at the border ($k=\pi/a$) of the Brillouin zone for poly-anthracene (n=3), poly-pentacene (n=5) and poly-heptacene (n=7) obtained via the CAMB3LYP~\cite{YanaiCPL2004,TawadaJCP2004} hybrid functional.
   }
\label{fig:HOMO_LUMO}
\end{figure}

\section{Screening due to environment}

If we compare the evGW estimate of the single-particle band-gap with the experimental values there seems to be a substantial difference. However, our computations are performed in vacuum while from the experimental point of view these polymers are deposited on Au(111) surfaces, which acts as a polarization environment, screening long-range interactions and strongly modifying correlation energies~\cite{NeatonPRL2006}. 

Therefore, we are not taking into account the effects of the environment on the quasiparticle properties. Indeed it has been shown~\cite{AmyOrg2005,NeatonPRL2006} in the case of anthracene and pentacene molecules that, when deposited on metallic surfaces, their quasi-particle band-gap was subjected to a renormalization between $~50\%$ and $~80\%$ with respect to the gas phase. If we apply such renormalization to our evGW values we find a quasiparticle band-gap between  $\sim2.66$ eV and $\sim1.67$ eV for poly-anthracene ($n=3$) and $\sim1.22$ eV and $\sim0.77$ eV for poly-pentacene ($n=5$), which are closer to the experimental results. The final discrepancy with respect to the experimental band-gap might be related to anharmonic effects, which can further reduce the energy difference between the HOMO and LUMO.

Notice however that neither in Ref.~\cite{CireraNatNanotec2020} nor in this work anharmonic quantum fluctuations are taken into account: these can play an important role in geometry optimization as well as in the Helmoltz free energy~\cite{MonacelliJCondMat,RomaninJPCL2021}, especially close to the quantum critical point.

\begin{table}[t]
    \centering
\begin{tabular}{ |c|c|c|}
 \hline
 n & $\Delta$E [eV]  & $\Delta$E [eV] \\
   & {\scriptsize (Experimental~\cite{CireraNatNanotec2020})} &  {\scriptsize (evGW)}  \\
 \hline
 3 & 1.50 & 3.33 \\
 \hline
 5 & 0.35 & 1.53 \\
 \hline
 7 & x & 2.78 \\
 \hline
\end{tabular}
\caption{Quasiparticle bandgap ($\Delta E$) for  poly-anthracene (n=3), poly-pentacene (n=5) and poly-heptacene (n=7). Experimental (Exp.) are taken from Ref.~\cite{CireraNatNanotec2020}while evGW refer to this work.}
\label{table:1}
\end{table}

\newpage

\section{Modulus Square of the Exciton's wavefunctions at\\ $q=0$}
\subsection{$n=5$ - Singlet}
\begin{figure}[H]
 \centering\includegraphics[height=0.8\textheight,keepaspectratio]{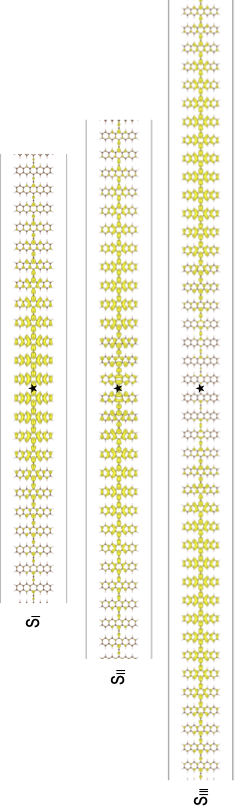}
 \caption
 {Modulus squared of the singlet exciton wavefunctions at $q=0$ for polypentacene ($n=5$). The black star represent the position of the hole.
   }
\label{fig:n5_singlet}
\end{figure}
\subsection{$n=5$ - Triplet}
\begin{figure}[H]
 \centering\includegraphics[height=0.8\textheight,keepaspectratio]{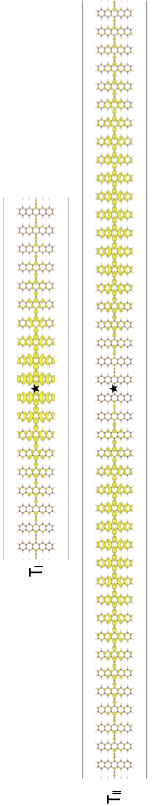}
 \caption
 {Modulus squared of the triplet exciton wavefunctions at $q=0$ for polypentacene ($n=5$). The black star represent the position of the hole.
   }
\label{fig:n5_triplet}
\end{figure}
\subsection{$n=7$ - Singlet}
\begin{figure}[H]
 \centering\includegraphics[width=0.7\linewidth]{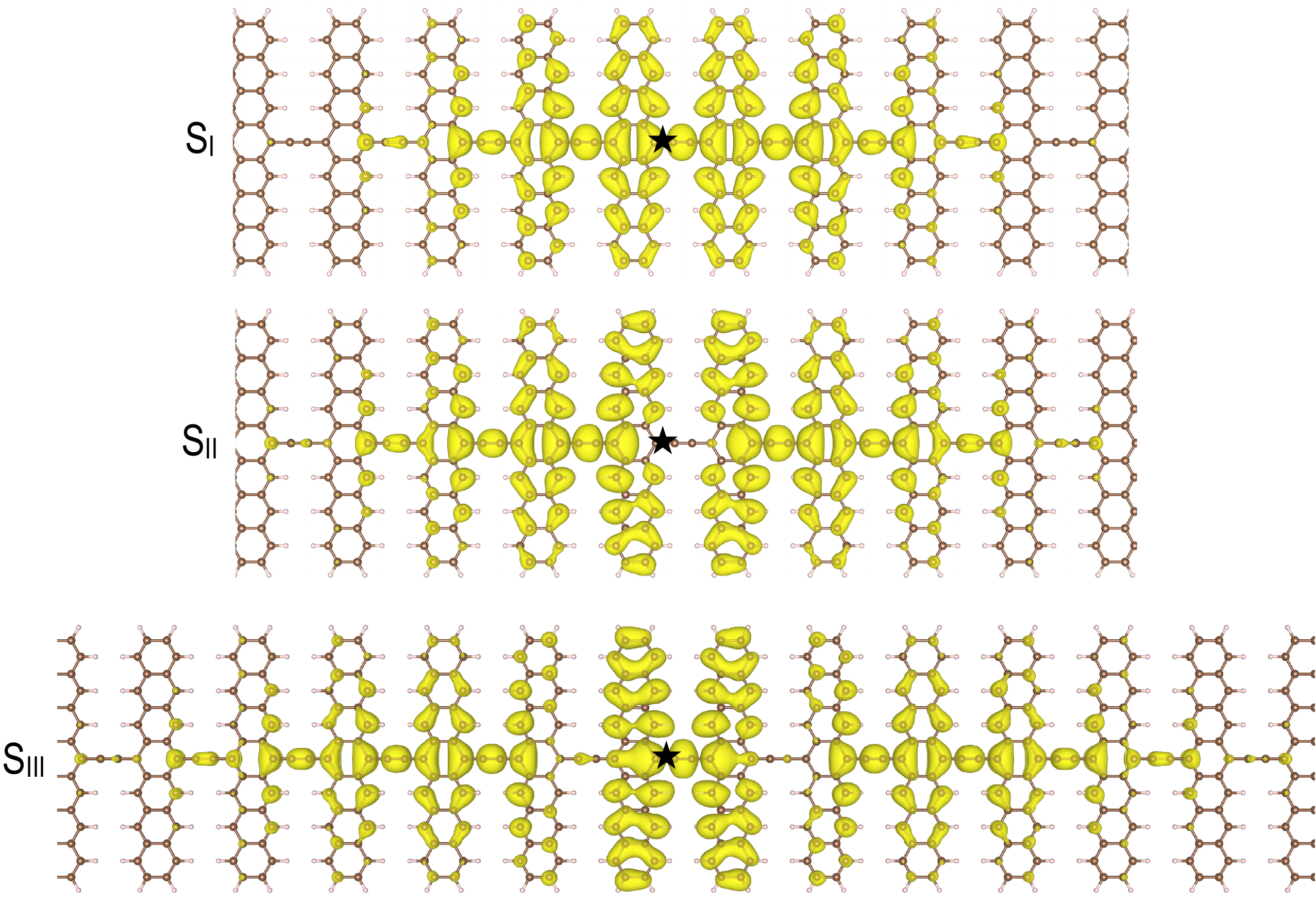}
 \caption
 {Modulus squared of the singlet exciton wavefunctions at $q=0$ for polyheptacene ($n=7$). The black star represent the position of the hole.
   }
\label{fig:n7_singlet}
\end{figure}
\subsection{$n=7$ - Triplet}
\begin{figure}[H]
 \centering\includegraphics[width=0.7\linewidth]{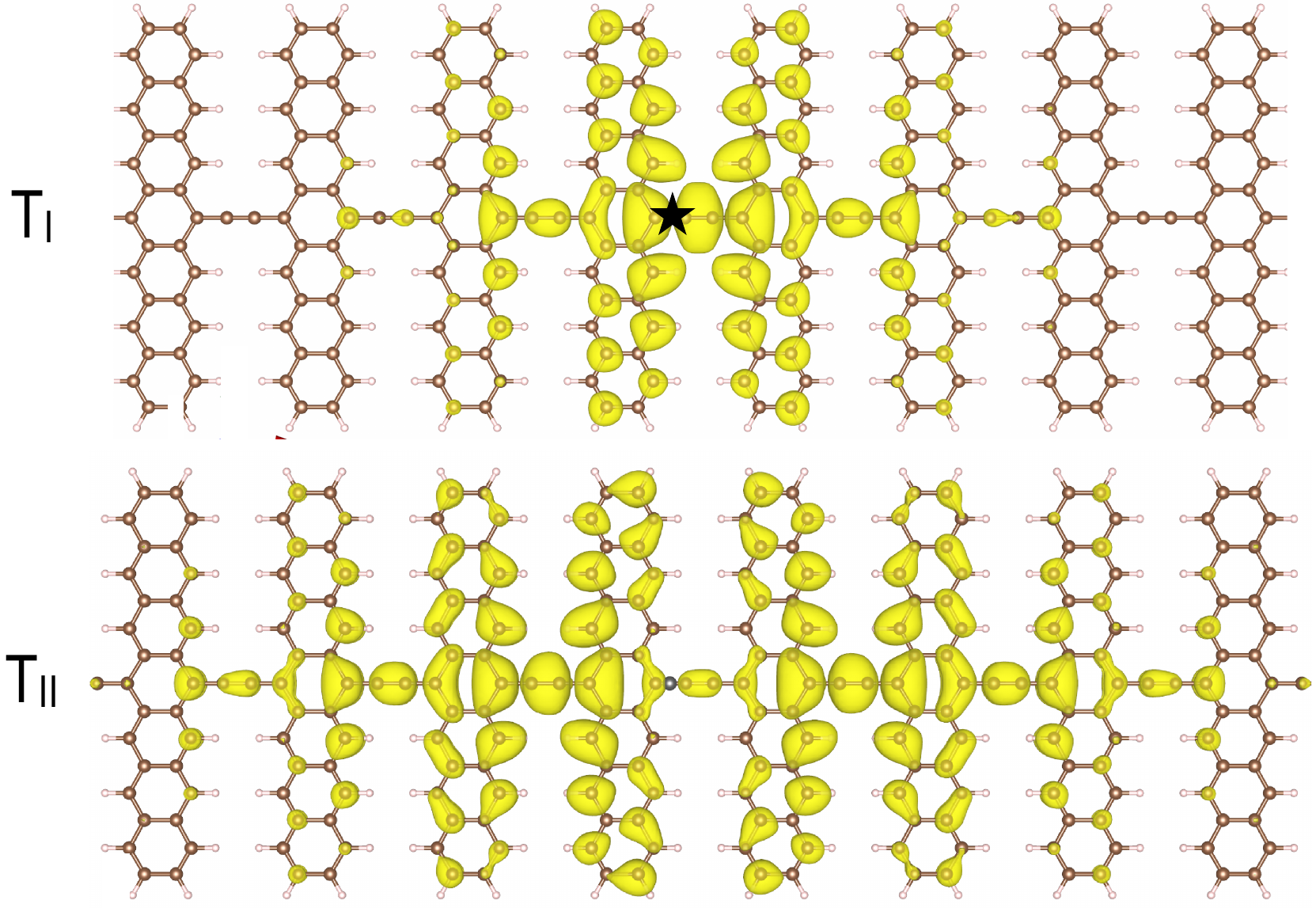}
 \caption
 {Modulus squared of the triplet exciton wavefunctions at $q=0$ for polyheptacene ($n=7$). The black star represent the position of the hole.
   }
\label{fig:n7_triplet}
\end{figure}

\newpage

\section{Magnetic groundstate}

As recently suggested in Ref.~\cite{Casula2022AFM} by means of Quantum Monte Carlo simulations, an infinite linear acene structure might develop a ground state  which is made of localized $\pi$ electrons whose spins are antiferromagnetically (AFM) ordered.

\begin{figure}[H]
 \centering\includegraphics[height=0.25\textheight,keepaspectratio]{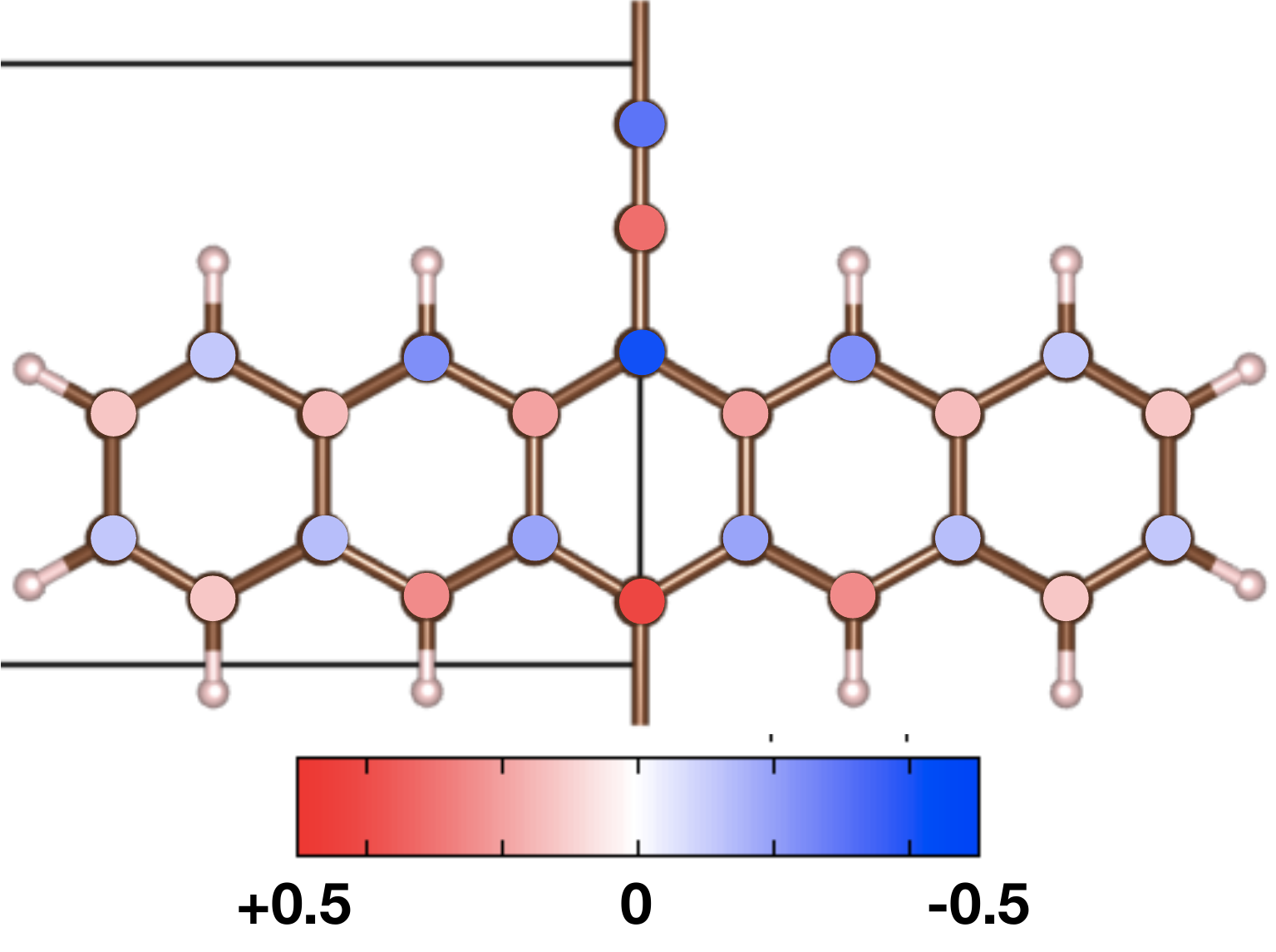}
 \caption
 {Stick-and-ball model of AFM poly-pentacene (n=5) obtained via the CAMB3LYP~\cite{YanaiCPL2004,TawadaJCP2004} hybrid functional. The atomic magnetic moments are plotted as color codes. The black horizontal lines indicate the unit cell.
   }
\label{fig:AFM_n5}
\end{figure}

\begin{figure}[H]
 \centering\includegraphics[height=0.25\textheight,keepaspectratio]{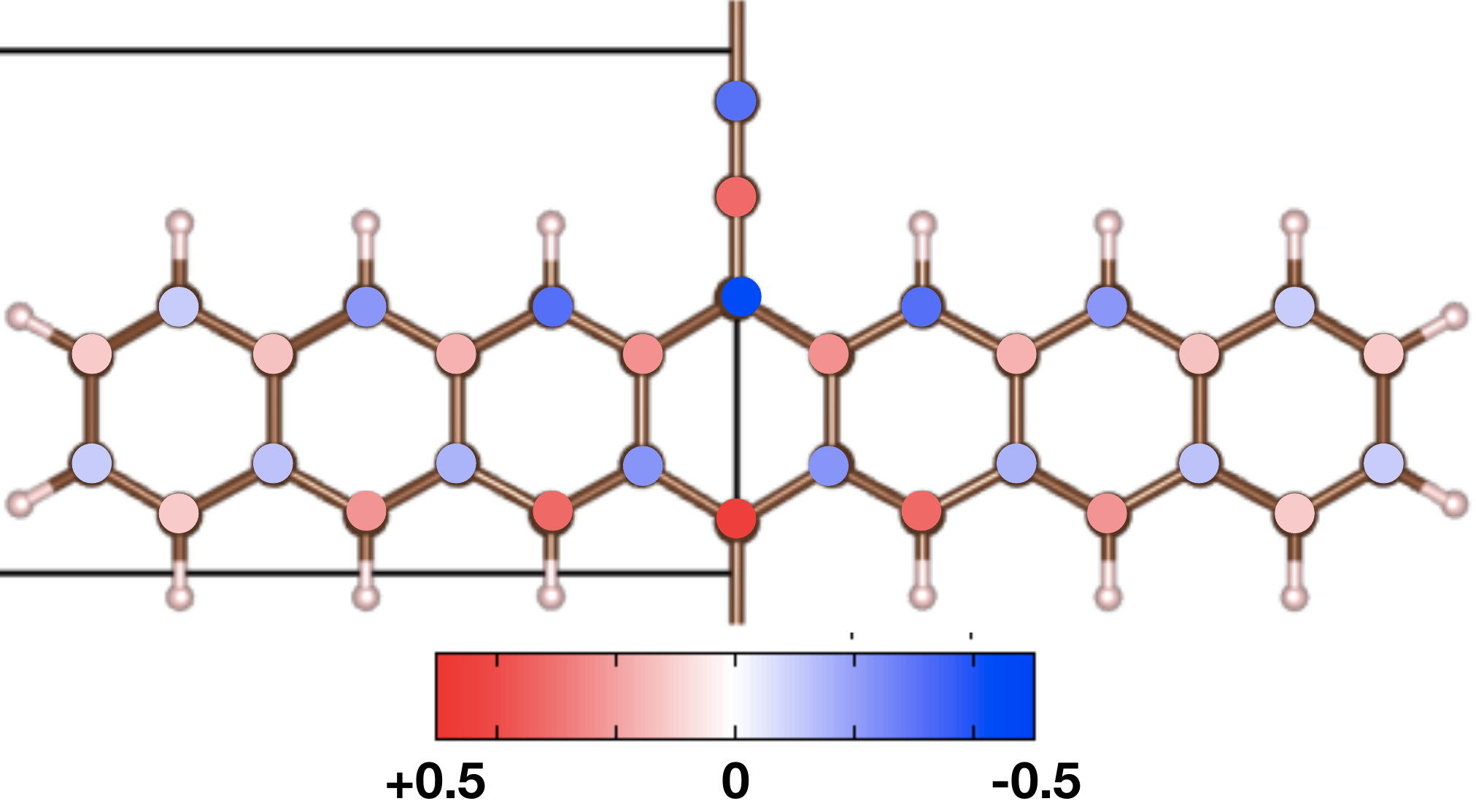}
 \caption
 {Stick-and-ball model of AFM poly-heptacene (n=7) obtained with the CAMB3LYP functional ~\cite{YanaiCPL2004,TawadaJCP2004}. The atomic magnetic moments are plotted as color codes. The black horizontal lines show the limits of the one dimensional unit cell.
   }
\label{fig:AFM_n7}
\end{figure}

We therefore performed spin-dependent DFT computation with the CAMB3LYP hybrid functional and found out that while poly-anthracene (n=3) remains paramagnetic (we did not  stabilize a magnetic order), poly-pentacene (n=5) and poly-heptacene (n=7) develop an AFM groundstate (see Fig.~\ref{fig:AFM_n5})-~\ref{fig:AFM_n7}). However, at odd with Ref. \cite{Casula2022AFM}, in our case the strongest magnetization is present along the principal axis of the polymers (i.e. on the ethynilene bridge), while it becomes smaller and smaller on the monomer unit as we get far from the central carbon ring. Moreover, the energy gain per carbon atom of the AFM phase with respect to the paramagnetic one is highest for n=5 (which is close to the quantum critical point) while it decreases when we pass to n=7 (see Tab.~\ref{table:3}). This suggests that for larger $n$ the ground state could become again non-magnetic. The study of larger values of $n$ is not relevant for the current work that focus on the optical properties across the topological transition. Not surprisingly, given the presence of the ethynilene bridge and the 1D periodicity along the bridge, the magnetic ordering is completely different of what happens in nanoribbons in vacuum Ref.~~\cite{Casula2022AFM}.

\begin{table}[H]
\begin{tabular}{ |c|c|c|c|c|c|c|c|}
 \hline
 n & c [$\AA$] & c [$\AA$] &  BLA [$\AA$]  & BLA [$\AA$] & $\Delta$E [eV]  & $\Delta$E [eV] & (E$_M$-E$_P$)/C \\
   & {\scriptsize (Paramagnetic)} &  {\scriptsize (Magnetic)}           &   {\scriptsize (Paramagnetic)}                & {\scriptsize (Magnetic)}           &   {\scriptsize (Paramagnetic)}      &   {\scriptsize (Magnetic)}       & {\scriptsize [meV/atom]}  \\
 \hline
 3 & 6.89 & 6.89 &  0.215 & 0.215 & 3.48 & 3.48 &  \\
 \hline
 5 & 6.92 & 6.90 & 0.210 & 0.160 & 2.35 & 3.50 & -4.8 \\
 \hline
 7 & 6.91 & 6.93 &  0.098 & 0.160 & 3.31 & 3.88 & -2.7 \\
 \hline
\end{tabular}
\caption{Comparison  between the paramagnetic (P) and the magnetic (M) groundstates for each polymer (n): lattice parameters (c), bond-length alternation (BLA) and single-particle bandgap ($\Delta$E). The last coloumn indicate the difference between the magnetic ($E_M$) and paramagnetic ($E_P$) total energies per carbon atom (C).}
\label{table:3}
\end{table}

\begin{figure}[H]
 \centering\includegraphics[height=0.25\textheight,keepaspectratio]{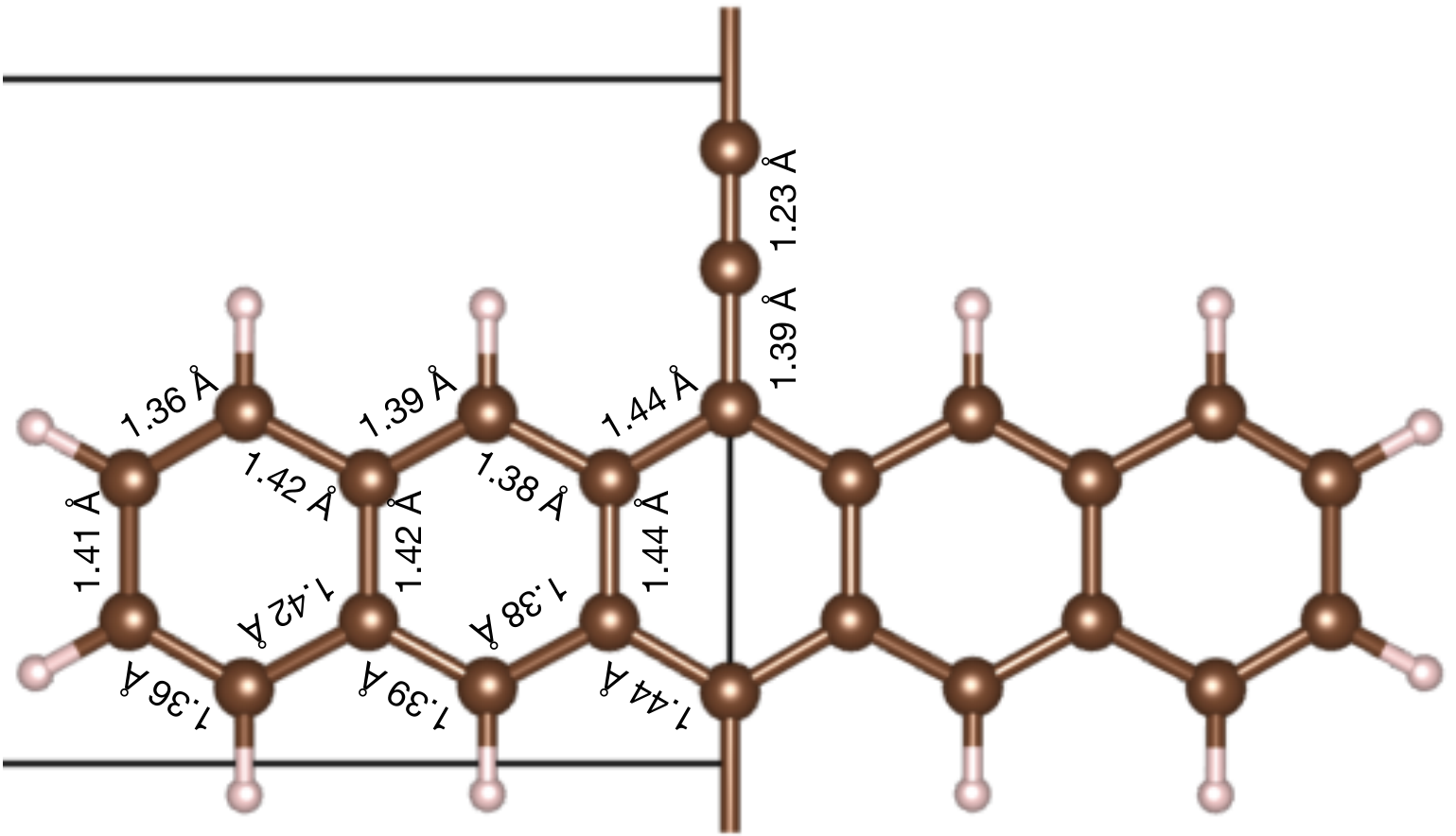}
 \caption
 {Stick-and-ball model of poly-pentacene (n=5) and bond legths obtained via the CAMB3LYP~\cite{YanaiCPL2004,TawadaJCP2004} hybrid functional.  Bond-lengths are written only for the left part of the polymer since it is symmetric with respect to its reciprocal axis. The black horizontal lines show the limits of the one dimensional unit cell. 
   }
\label{fig:magnetic_structure_n5}
\end{figure}

\begin{figure}[H]
 \centering\includegraphics[height=0.25\textheight,keepaspectratio]{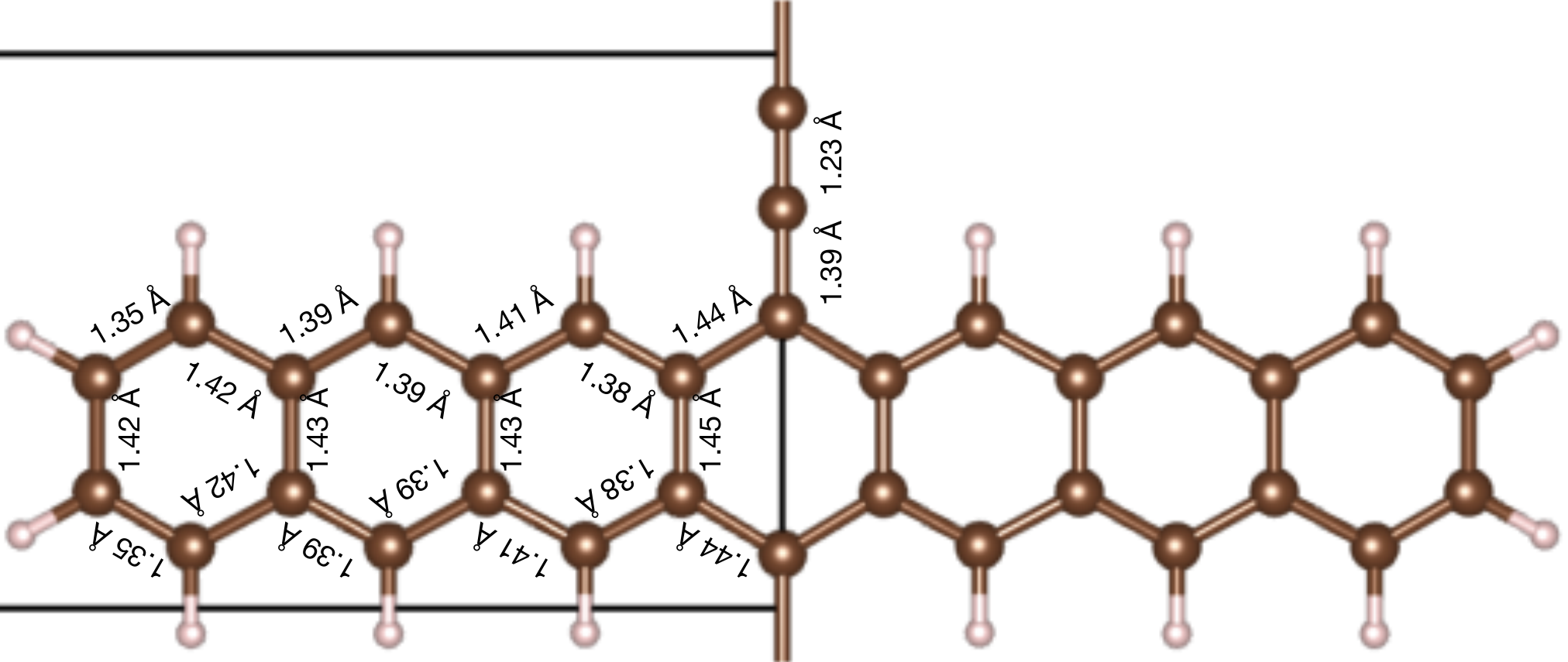}
 \caption
 {Stick-and-ball model of poly-heptacene (n=7) with relative bond lengths. The magnetic groundstate geometry has been obtained via the CAMB3LYP~\cite{YanaiCPL2004,TawadaJCP2004} hybrid functional. The black horizontal lines indicate the unit cell. Bond-lengths are written only for the left part of the polymer since it is symmetric with respect to its reciprocal axis.
   }
\label{fig:magnetic_structure_n7}
\end{figure}

In the AFM groundstate, the exchange of the HOMO/LUMO levels already appear between $n=3$ and $n=5$ (Fig.~\ref{fig:HOMO_LUMO_AFM}) with a reduction of the BLA of just 0.05~\AA~and an increment of the quasi-particle bandgap (see Fig.~\ref{fig:magnetic_structure_n5}-~\ref{fig:magnetic_structure_n7} and Tab.~\ref{table:3}). 

\begin{figure}[H]
 \centering\includegraphics[width=1.00\linewidth]{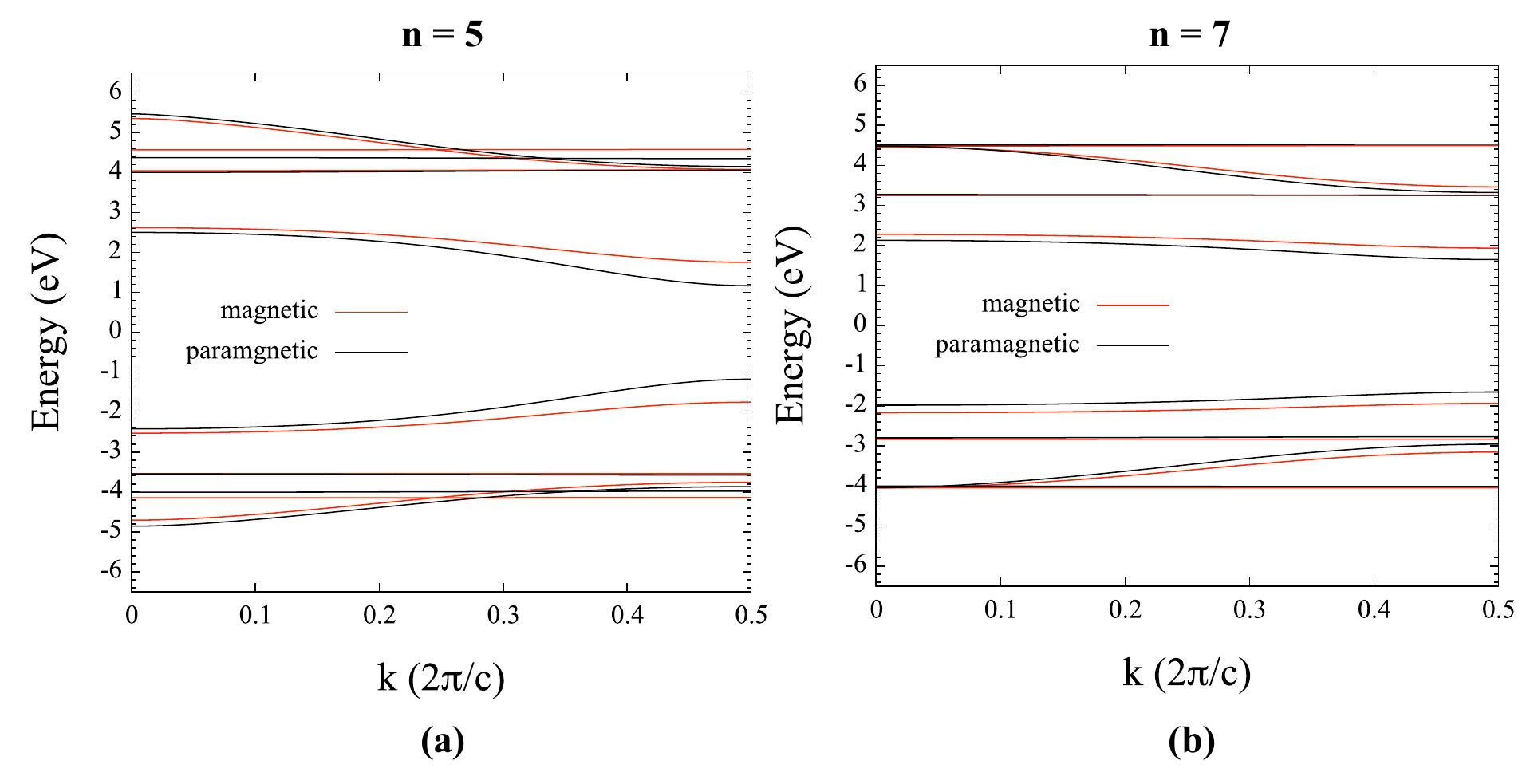}
 \caption
 {Electronic bandstructure for poly-pentacene (n=5) and poly-heptacene (n=7) in their magnetic groundstate obtained via the CAMB3LYP~\cite{YanaiCPL2004,TawadaJCP2004} hybrid functional.
   }
\label{fig:Magnetic_bands}
\end{figure}

\begin{figure}[H]
 \centering\includegraphics[width=0.75\linewidth]{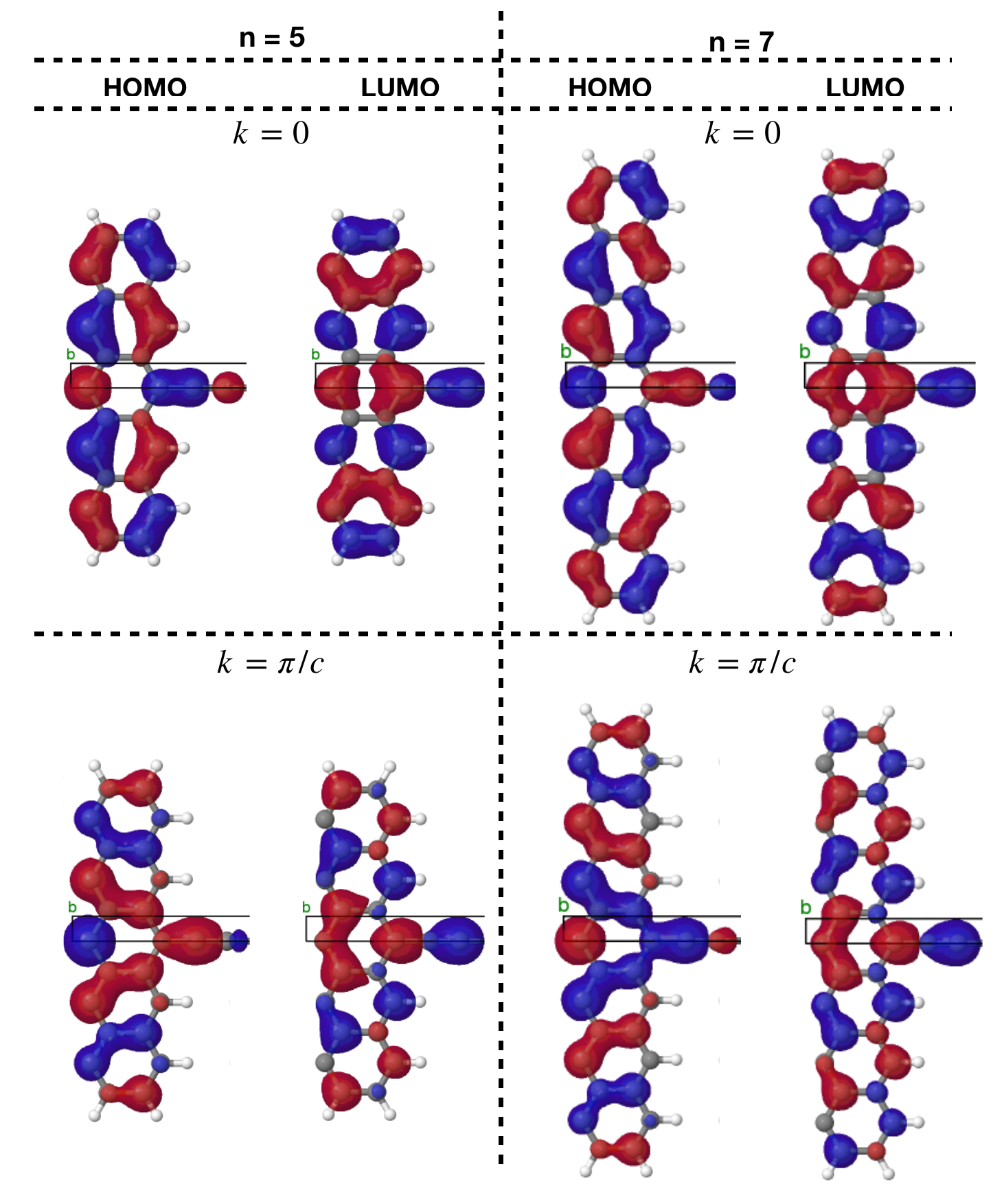}
 \caption
 {HOMO/LUMO wavefunctions at the center ($k=0$) and at the border ($k=\pi/a$) of the Brillouin zone for poly-pentacene (n=5) and poly-heptacene (n=7) in their AFM groundstate obtained via the CAMB3LYP~\cite{YanaiCPL2004,TawadaJCP2004} hybrid functional.
   }
\label{fig:HOMO_LUMO_AFM}
\end{figure}

From a qualitative and quantitative point-of-view the electronic band structures of poly-pentacene and poly-heptacene are almost unaffected  by magnetism, as shown in Fig. \ref{fig:Magnetic_bands}, except for a weak increment of the quasiparticle bandgap and the exchange of the HOMO/LUMO levels for n=5 (cf. Fig.~\ref{fig:CAMB3LYP_bands} and Fig.~\ref{fig:Magnetic_bands}). As the system is antiferromagnetic, the minority and majority spin bands are degenerate and we do not expect qualitative differences  from the non-magnetic state for what concerns
the optical absorption, except for minor corrections to the magnitude of the excitonic binding energies, optical gaps and singlet-triplet splittings.\\

We would like, however, to point out that a magnetic groundstate is unlikely to occu at room temperature in supported samples (while we cannot exclude its stability at low temperatures in vacuum).  Beside the fact that the nature of magnetism in this compound is along its one-dimensional periodic direction and quantum fluctuations could play an important role in suppressing the order,  from the experimental point of view, no signatures of a magnetic groundstate are reported~\cite{CireraNatNanotec2020,GonzalezCondMat2021}. This can be due to the general weakness of magnetism in acenes and in nanoribbons that is quickly suppressed by either the hybridisation with the substrate or weak intrinsic doping 
(see \cite{Blackwell2021} and references therein).

\newpage

\bibliography{reference_SI}

\end{document}